\documentclass[11pt]{article}
\usepackage{graphicx}

\title{Bound States and Critical Behavior of the Yukawa Potential\thanks{LXQ is supported
by the Key Project of National Science Foundation (10235040), Key
Project of National Ministry of Eduction (105135), Project of the
Chinese Academy of Sciences (KJCX2-SW-N10) and Guangdong Ministry
of Education. HK is supported by NSERC Canada.} }

\author{Xiang-Qian Luo\thanks{Corresponding author.
Email: stslxq@zsu.edu.cn}, Yong-Yao Li\\
{\small\sl Department of Physics, Zhongshan University,
Guangzhou 510275, China}\\
Helmut Kr\"oger\\
{\small\sl D\'epartement de Physique, Universit\'e Laval, Qu\'ebec G1K
7P4, Canada}\\
}

\date{\today}

\begin{document}
\maketitle

\begin{abstract}
We investigate the bound states of the Yukawa
potential $V(r)=-\lambda \exp(-\alpha r)/ r$,
using different algorithms: solving the
Schr\"odinger equation numerically and our Monte Carlo Hamiltonian approach.
There is a critical $\alpha=\alpha_C$, above which no bound state exists.
We study the relation between $\alpha_C$ and $\lambda$ for
various angular momentum quantum number $l$, and find in atomic units,
$\alpha_{C}(l)=
\lambda [A_{1} \exp(-l/ B_{1})+ A_{2} \exp(-l/ B_{2})]$,
with $A_1=1.020(18)$, $B_1=0.443(14)$, $A_2=0.170(17)$, and $B_2=2.490(180)$.
\end{abstract}

\leftline{{\bf Key words:} Yukawa potential, bound states, critical behavior.}

%\leftline{{\bf PACS numbers:} 02.50.Ng, 03.65.-w, 03.65.Ge}

\maketitle

%\setcounter{page}{0}
%\newpage

\section{Introduction}
\label{SEC1}

About 70 years ago,
Yukawa proposed  his famous meson theory
to describe the interactions between nucleons\cite{Yukawa:1935xg}, in which
the potential is given by
\begin{eqnarray}
V(r)=-\lambda {\exp(-\alpha r)\over r} .
\label{eq1}
\end{eqnarray}
Here $\lambda$ and $1/\alpha$ are respectively the strength and
range of the nucleon force. Last few decades have seen a lot of
ongoing analytical and numerical
efforts\cite{Sachs,Rogers,McEnnan,Gerry,Kroger1,Girard:1987xj,GARAVELLI,Gomes,Yukalov:1998gk,Brau}
on exploring the properties of the Yukawa potential. Up to very
recently, this model still receives great
attention\cite{Bertini,Dean,DeLeo:2004vy,Khrapak,Iorio:2002jy},
for it plays an important role not only in particle/nuclear
physics, but also in many other branches: atomic physics, chemical
physics, gravitational plasma physics, and solid-state physics. In
plasma physics, it is known as the Debye-H\"uckel potential; while
in solid-state physics and atomic physics, it is named as the
Thomas-Fermi or screened Coulomb potential.

Critical phenomena exist not only in Quantum Chromodynamics (QCD)
at finite temperature/chemical potential\cite{Gregory:1999pm,Luo:2000xi,Fang:2002rk,Luo:2004mc}, but also in
the Yukawa potential in Quantum Mechanics (QM)
\cite{Harris,Schey,Rogers,GARAVELLI,Gomes,Yukalov:1998gk,Brau}.
For $\alpha=0$, the Yukawa potential reduces to the Coulomb potential, and
it is known to have infinite number of bound states.
For $\alpha=\infty$, there is no interaction at all, and the system is completely free.
Therefore, when $\alpha$ is non-zero,
the Yukawa potential has some features very different from the Coulomb potential.
One expects that the number of bound states is limited, because the interactions are screened. When $\alpha$ is sufficiently large, one expects that the bound states disappear.

In this paper, we study systematically
the bound states and the critical behavior of the Yukawa
potential, using different algorithms: (1)
numerical solution to differential equation\cite{NR};
(2) Monte Carlo Hamiltonian approach\cite{Jirari:1999jn},
developed by some authors of the present article.
The effects of the input parameters on the wave function and energy,
and therefore on the critical behavior of the system are investigated in great details.
We also propose a general relation among the critical $\alpha$, $\lambda$
and the angular momentum $l$.

The rest of paper is organized as follows. In Sec. \ref{SEC2}, we describe briefly the
numerical methods.
In Sec. \ref{SEC3}, we show the results.
The critical behavior is discussed in Sec. \ref{SEC4} and the discussions are given in Sec. \ref{SEC5}

\section{Algorithms}
\label{SEC2}

\subsection{Schr\"odinger equation}

Because there is a spherical symmetry in the Yukawa potential Eq.
(\ref{eq1}), the Schr\"odinger equation is reduced to a radial one
\begin{eqnarray}
{d^2 u(r) \over dr^2} + \left[ {2m \over \hbar^2} \left(E- U \left(r \right) \right ) \right] u(r)=0,
\label{eq2}
\end{eqnarray}
where $u(r)$ is the reduced wave function, and $U(r)$ is a $l$ dependent
potential
\begin{eqnarray}
U(r)=V(r)+{l(l+1)\hbar^2\over 2mr^2} .
\label{eq3}
\end{eqnarray}

Runge-Kutta and Numerov algorithms are two widely employed methods\cite{NR}
for numerically solving such a Sturm-Liouville problem. Let us take the
Numerov algorithm as an example.
With $q(r)= \left(E- U \left(r \right) \right )2m/ \hbar^2$,
the Schr\"odinger equation
\begin{eqnarray}
{d^2 u(r) \over dr^2} + q(r) u(r)=0
\label{eq4}
\end{eqnarray}
is differentiated as
\begin{eqnarray}
c_{n+1} u_{n+1} + c_{n-1} u_{n-1} = c_n u_n + O(\epsilon^6),
\label{eq5}
\end{eqnarray}
where $\epsilon$ is the integration step, and
\begin{eqnarray}
c_{n+1} &=& 1+{\epsilon^2 \over 12} q_{n+1} ,
\nonumber \\
c_{n} &=& 2-{5\epsilon^2 \over 6} q_{n} ,
\nonumber \\
c_{n-1} &=& 1+{\epsilon^2 \over 12} q_{n-1} .
\label{eq6}
\end{eqnarray}
The boundary condition is $u(0)=u(r_{max})=0$.
One can carry out the integration from $r=0$ and $r=r_{max}$, and match the wave function and the its derivative
in the potential well region.
Then we can combine the Sturm theorem and bisection method
to find the eigenvalues and eigen-functions.

\subsection{Monte Carlo Hamiltonian}

Although most 1D QM problems can well be studied by
Runge-Kutta and Numerov algorithms, it is impossible to extend them to
high dimensional QM or quantum field theory.
Monte Carlo (MC) method with importance sampling\cite{Metropolis:1953am} is able to compute
high dimensional (and even ``infinite" dimensional)
path integrals.
Unfortunately, using the standard Lagrangian MC technique,
it is extremely difficult to estimate wave functions and
spectrum of excited states.
Let us take as an example a new type of hadrons predicted by QCD,
the so-called glueballs\cite{Luo:1996ha,Luo:96}, which are bound states of gluons.
Wave functions in conjunction with the energy spectrum
contain more physical information on glueballs.

Some years ago, we proposed a new approach\cite{Jirari:1999jn}
(named Monte Carlo Hamiltonian method or MCH) to investigate this problem.
A lot of models\cite{Jirari:1999jn,Luo:1999dz,Luo:1999yi,Jirari:1999ie,Jiang:2000bs,Huang:2000hb,Luo:2001fr,Luo:2002rx}
in QM have been used to test the method.
This method has also been applied
to scalar field theories\cite{Luo:2001gb,Huang:1999fn,Kroger:2002pb,Kroger:2002rh}.

We briefly mention the basic idea of MC Hamiltonian\cite{Jirari:1999jn}.
Starting from the Euclidean action $S_{E} = \int dt ~  \left( m \dot{r}^{2}/2 + U(r) \right) $,
we consider the transition amplitude in imaginary time
between time $t=0$ and $t=T$ for all combinations of positions
$r_{i}, r_{j} \in \{r_{1},\dots,r_{N}\}$,
\begin{eqnarray}
M_{i,j}(T)
= <r_{i} | e^{-H T/\hbar} | r_{j}>
= \int [dr] \exp[ -S_{E}[r]/\hbar ]\bigg |_{r_{j},0}^{r_{i},T} ,
\label{eq:DefHeff}
\end{eqnarray}
which is a path integral\cite{Feynman} and can be computed
by MC simulation with importance sampling\cite{Metropolis:1953am},
at least in QM and the scalar models.
The $N \times N$ transition matrix $M$
can be diagonalized by a unitary transformation
and used to construct the effective Hamiltonian\cite{Jirari:1999jn}.
Then we obtain the eigenvalues and eigen-functions of the effective Hamiltonian.

Although MCH has been very successful applied to many local
potentials, a special treatment has to be taken for QM with
potentials having singularity at $r=0$. The trick is to replace
$V(r)$ by
\begin{eqnarray}
V_{eff}(r) ={{\exp(2r/R)-1}\over
{\exp(2r/R)+1}}V(r) .
\end{eqnarray}
$V_{eff} (r) $ is equal to  $V(r)$ in the $R \to 0$ limit.
We first compute the spectrum and wave functions at some finite $R$,
and then extrapolate the
data to $R \to 0$. In Ref. \cite{Luo:2003xr}, we successfully applied this method
to the hydrogen system.

\section{Numerical results}
\label{SEC3}

\begin{figure} [htbp]
\begin{center}
\includegraphics[totalheight=3.0in]{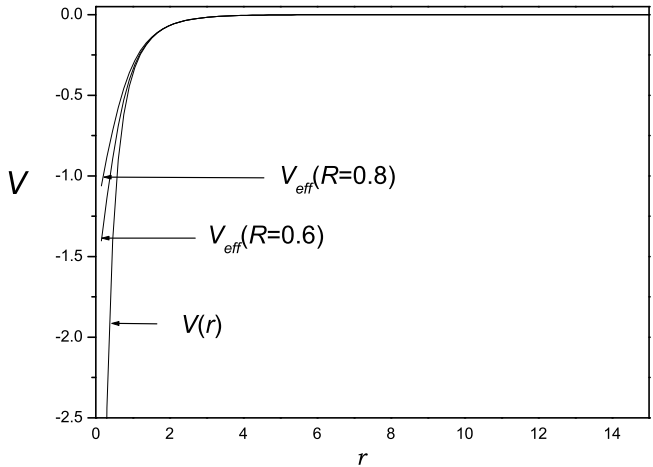}
\end{center}
\caption{$V_{eff}(r)$ as a function of $r$.}
\label{fig1}
\end{figure}

Both the Numerov and Runge-Kutta algorithms
can be directly applied to the Yukawa potential with very high precision.
For MCH, as mentioned in Sec. \ref{SEC2},
we have to introduce a cutoff $R$ for $V_{eff}(r)$.
Figure \ref{fig1} shows the dependence of $V_{eff}(r)$ on $R$.
For computing the transition matrix elements Eq. (\ref{eq:DefHeff}),
we choose  the regular basis:
\begin{eqnarray}
r_{j}=j \times \Delta r, & &  j=1, 2, \ldots, N ,
\end{eqnarray}
and let $\hbar=1$ , $m=1$,  $\Delta r=1.0$, and $N=100$.

\begin{figure} [htbp]
\begin{center}
\includegraphics[totalheight=3.0in]{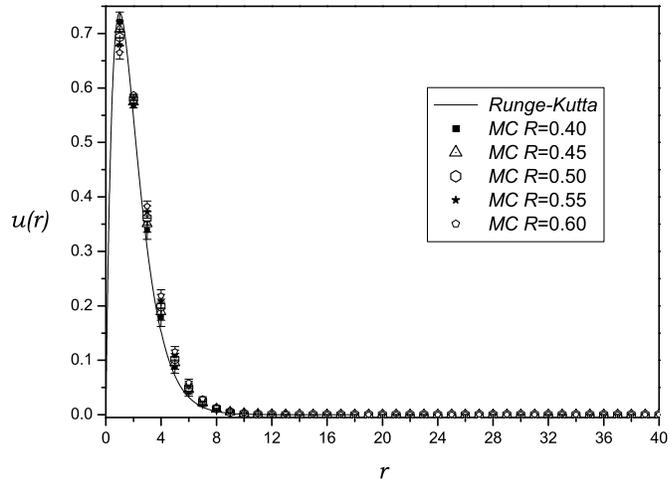}
\end{center}
\caption{Reduced wave function for the ground state
at $l=0$, and $\lambda=1$ and $\alpha=0.1$.}
\label{fig2}
\end{figure}

\begin{figure} [htbp]
\begin{center}
\includegraphics[totalheight=3.0in]{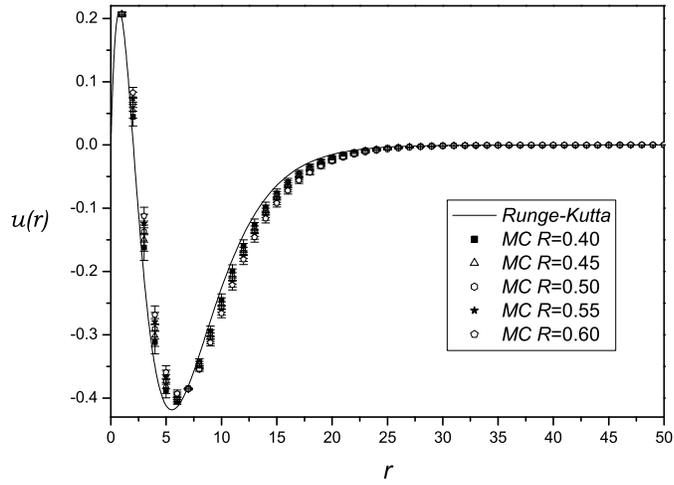}
\end{center}
\caption{Same as Fig. \ref{fig2}, but for the first excited state.}
\label{fig3}
\end{figure}

\begin{figure} [htbp]
\begin{center}
\includegraphics[totalheight=3.0in]{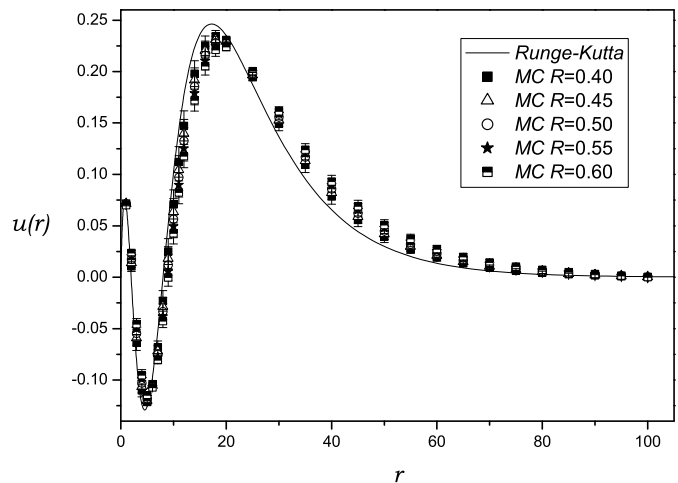}
\end{center}
\caption{Same as Fig. \ref{fig2}, but for the second excited state.}
\label{fig4}
\end{figure}

Figures \ref{fig2}-\ref{fig4} show the reduced wave function of
the first three bound states at $l=0$, $\lambda=1$ and
$\alpha=0.1$. These figures indicate that when $R \to 0$ the MCH
data approach the Runge-Kutta/Numerov data. Table \ref{tab1} lists
some spectrum data for $l=0$, $\lambda=1$ and $\alpha=0.1$, 0.2,
0.3, 0.4. The error analysis follows the method described in Ref.
\cite{Luo:2003xr}. Within error bars, the MCH data agree with
those obtained by the Runge-Kutta/Numerov algorithms. We have also
confirmed that for sufficient large $r_{max}$, which is the limit
of integration, the results are stable and independent on
$r_{max}$.

\begin{table}
\begin{center}
\begin{tabular}{|c|c|c|c|}\hline
$\alpha$  & Quantum number $n$ & Runge-Kutta/Numerov & MCH ($R\to 0$) \\ \hline
0.1 & 1& -0.4070 & -0.4051 $\pm$ 0.0612 \\  \cline{2-4}
& 2 & -0.0499 & -0.0463 $\pm$ 0.0069 \\ \cline{2-4}
& 3 &-0.0032 & -0.0031 $\pm$ 0.0008 \\ \hline
0.2 & 1 & -0.3268 & -0.3345 $\pm$ 0.0604 \\ \cline{2-4}
&  2 & -0.0121 & -0.0120 $\pm$ 0.0036 \\ \hline
0.3 & 1 & -0.2576 & -0.2453 $\pm$ 0.0584 \\ \hline
0.4 & 1 & -0.1984 & -0.2069 $\pm$ 0.0545 \\ \hline
\end{tabular}
\end{center}
\caption{Comparison of  some eigenvalues for $l=0$, $\lambda=1$, and some $\alpha$,
between the Runge-Kutta/Numerov and MCH methods, where the MCH
data have been extrapolated to the $R=0$ limit.}
\label{tab1}
\end{table}

\section{Critical Phenomena}
\label{SEC4}

\begin{figure} [htbp]
\begin{center}
% GNUPLOT: LaTeX picture with Postscript
\begingroup%
  \makeatletter%
  \newcommand{\GNUPLOTspecial}{%
    \@sanitize\catcode`\%=14\relax\special}%
  \setlength{\unitlength}{0.1bp}%
\begin{picture}(2519,1943)(0,0)%
\special{psfile=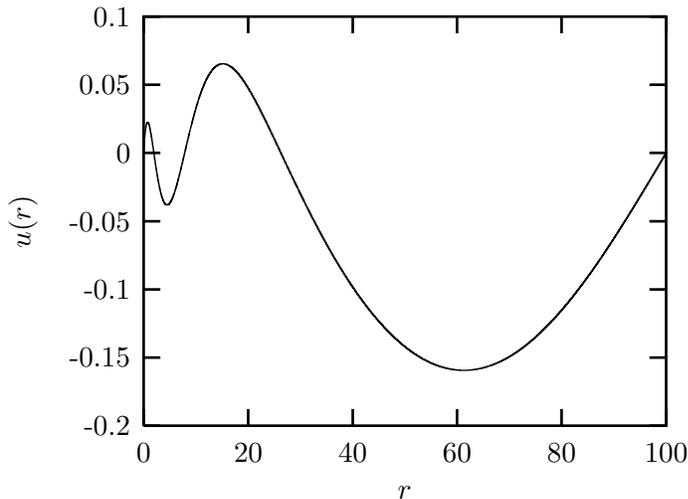 llx=0 lly=0 urx=504 ury=454 rwi=5040}
\put(1484,50){\makebox(0,0){$r$}}%
\put(100,1071){%
\special{ps: gsave currentpoint currentpoint translate
270 rotate neg exch neg exch translate}%
\makebox(0,0)[b]{\shortstack{$u(r)$}}%
\special{ps: currentpoint grestore moveto}%
}%
\put(2469,200){\makebox(0,0){100}}%
\put(2075,200){\makebox(0,0){80}}%
\put(1681,200){\makebox(0,0){60}}%
\put(1288,200){\makebox(0,0){40}}%
\put(894,200){\makebox(0,0){20}}%
\put(500,200){\makebox(0,0){0}}%
\put(450,1843){\makebox(0,0)[r]{0.1}}%
\put(450,1586){\makebox(0,0)[r]{0.05}}%
\put(450,1329){\makebox(0,0)[r]{0}}%
\put(450,1071){\makebox(0,0)[r]{-0.05}}%
\put(450,814){\makebox(0,0)[r]{-0.1}}%
\put(450,557){\makebox(0,0)[r]{-0.15}}%
\put(450,300){\makebox(0,0)[r]{-0.2}}%
\end{picture}%
\endgroup
 
\end{center}
\caption{Reduced wave function for the third excited state
at $l=0$, and $\lambda=1$ and $\alpha=0.1$, with $r_{max}=100$.}
\label{fig5}
\end{figure}

\begin{figure} [htbp]
\begin{center}
% GNUPLOT: LaTeX picture with Postscript
\begingroup%
  \makeatletter%
  \newcommand{\GNUPLOTspecial}{%
    \@sanitize\catcode`\%=14\relax\special}%
  \setlength{\unitlength}{0.1bp}%
\begin{picture}(2519,1943)(0,0)%
\special{psfile=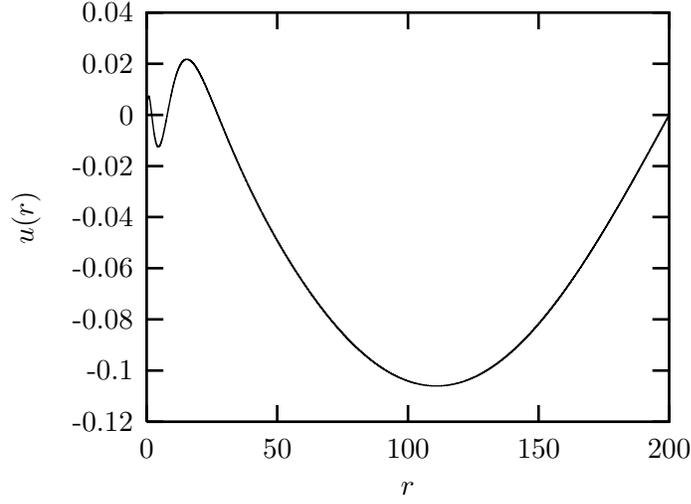 llx=0 lly=0 urx=504 ury=454 rwi=5040}
\put(1484,50){\makebox(0,0){$r$}}%
\put(100,1071){%
\special{ps: gsave currentpoint currentpoint translate
270 rotate neg exch neg exch translate}%
\makebox(0,0)[b]{\shortstack{$u(r)$}}%
\special{ps: currentpoint grestore moveto}%
}%
\put(2469,200){\makebox(0,0){200}}%
\put(1977,200){\makebox(0,0){150}}%
\put(1485,200){\makebox(0,0){100}}%
\put(992,200){\makebox(0,0){50}}%
\put(500,200){\makebox(0,0){0}}%
\put(450,1843){\makebox(0,0)[r]{0.04}}%
\put(450,1650){\makebox(0,0)[r]{0.02}}%
\put(450,1457){\makebox(0,0)[r]{0}}%
\put(450,1264){\makebox(0,0)[r]{-0.02}}%
\put(450,1072){\makebox(0,0)[r]{-0.04}}%
\put(450,879){\makebox(0,0)[r]{-0.06}}%
\put(450,686){\makebox(0,0)[r]{-0.08}}%
\put(450,493){\makebox(0,0)[r]{-0.1}}%
\put(450,300){\makebox(0,0)[r]{-0.12}}%
\end{picture}%
\endgroup
 
\end{center}
\caption{Same as Fig. \ref{fig5}, but with $r_{max}=200$.}
\label{fig6}
\end{figure}

As mentioned in Sec. \ref{SEC1},  for a finite $\alpha$, the
number of bound states is finite. Let us first take the system
with $l=0$, $\lambda=1$ and $\alpha=0.1$ as an example. Figures
\ref{fig5} and \ref{fig6} show the reduced wave function of the
third excited state with $r_{max}=100$ and $r_{max}=200$
respectively. The location of the lowest valley goes further and
further away from the origin, with increasing limit of integration
$r_{max}$. It will goes to $r=\infty$ when $r_{max} \to \infty$.
The eigenvalue $E$ for the third excited state is $8.1037 \times
10^{-4}$ for $r_{max}=100$, $1.5533 \times 10^{-4}$ for
$r_{max}=200$, and $6.4491 \times 10^{-5}$ for $r_{max}=300$.
I.e., $E$ goes to zero when $r_{max} \to \infty$. Therefore the
third excited state is no longer bounded. This also applies to
higher states.

\begin{table}
\begin{center}
\begin{tabular}{|c|c|c|c|c|c|}
\hline
~  & $\lambda=1$ & $\lambda=2$ & $\lambda=3$ & $\lambda=4$ & $\lambda=5$ \\ \hline
$\alpha=0$ & $\infty$ & $\infty$ & $\infty$ & $\infty$ & $\infty$ \\ \hline
$\alpha=0.1$ & 3 & 5 & 6 & 7 & 7\\ \hline
$\alpha=0.2$ & 3 & 3 & 4 & 5 & 5\\  \hline
$\alpha=0.3$ & 2 & 2 & 3 & 3 & 3 \\ \hline
$\alpha=0.4$ & 1 & 2 & 3 & 3 & 3 \\ \hline
$\alpha=0.5$ & 1 & 2 & 2 & 3 & 3 \\  \hline
$\alpha=0.6$ & 1 & 2 & 2 & 2 & 3 \\ \hline
$\alpha \in [0.7,0.9]$ & 1 & 1 & 2 & 2 & 2 \\  \hline
$\alpha \in [1.0,1.1]$ & 1 & 1 & 1 & 2 & 2 \\  \hline
$\alpha=1.2$ & - & 1 & 1 & 2 & 2 \\  \hline
$\alpha \in [1.3,1.4]$ & - & 1 & 1 & 1 & 2  \\ \hline
$\alpha \in [1.5,2.3]$ & - & 1 & 1 & 1 & 1 \\  \hline
$\alpha \in [2.4,3.5]$ & - & - & 1 & 1 & 1 \\  \hline
$\alpha \in [3.6,4.7]$ & -& - & - & 1 & 1 \\  \hline
$\alpha \in [4.8,5.9]$ & - & - & -& 1 & 1 \\ \hline
$\alpha=6.0$ & - & - & - & - & - \\  \hline
\end{tabular}
\end{center}
\caption{Number of bound states at $l=0$ for various $\lambda$ and
$\alpha$.}
\label{tab2}
\end{table}

Table \ref{tab2} shows the number of bound states for $l=0$
and various $\lambda$ and $\alpha$.
From the table, we find that when $\alpha$ is sufficiently large,
the attractive potential is so screened so that the system no longer has a bound state.
The critical value for $\alpha$ is
named critical screening parameter $\alpha_{C}$.
After fine scanning $\alpha$ and studying the dependence of $u(r)$ on $r_{max}$,
we obtain in Tab. \ref{tab3} the value of  $\alpha_C$
for $l=0$ and various $\lambda$. It is clear that $\alpha_{C}$ satisfy the
relation
\begin{eqnarray}
\alpha_{C}(l=0) = 1.1906 \lambda .
\end{eqnarray}
This tells us that the physics is the same for all $\lambda$, after rescaling the variables.

\begin{table}
\begin{center}
\begin{tabular}{|c|c|c|c|c|c|}\hline
  $\lambda$ & 1& 2 & 3 & 4 & 5 \\ \hline $\alpha_{C}$ & 1.1906 &
  2.3812 & 3.5718 & 4.7624  & 5.9530  \\ \hline
\end{tabular}
\end{center}
\caption{Dependence of $\alpha_{C}$ on $\lambda$ for $l=0$.}
\label{tab3}
\end{table}

We have obtained the result of Tab. \ref{tab3} by two criteria:
(1) There is a unstable peak (or valley) in the ground wave function, with its location
moving to $r \to \infty$ when  $r_{max} \to \infty$;
(2) The eigenvalue of the ground state becomes zero or positive as $r_{max} \to \infty$.

\begin{figure} [htbp]
\begin{center}
\includegraphics[totalheight=3.0in]{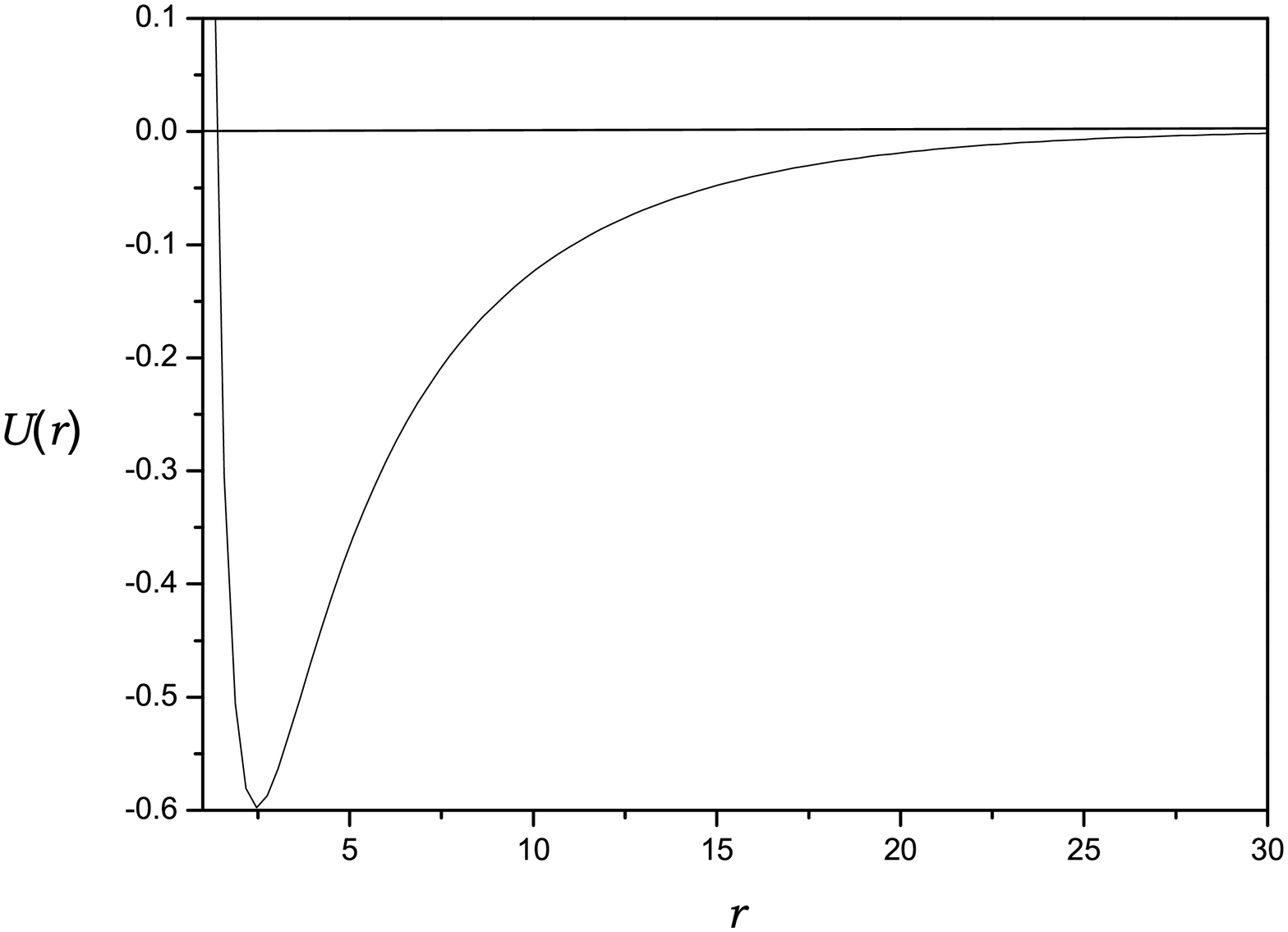}
\end{center}
\caption{$U(r)$ for $l \ne 0$ when $\alpha << \alpha_C$.}\label{fig7}
\end{figure}

\begin{figure} [htbp]
\begin{center}
\includegraphics[totalheight=3.0in]{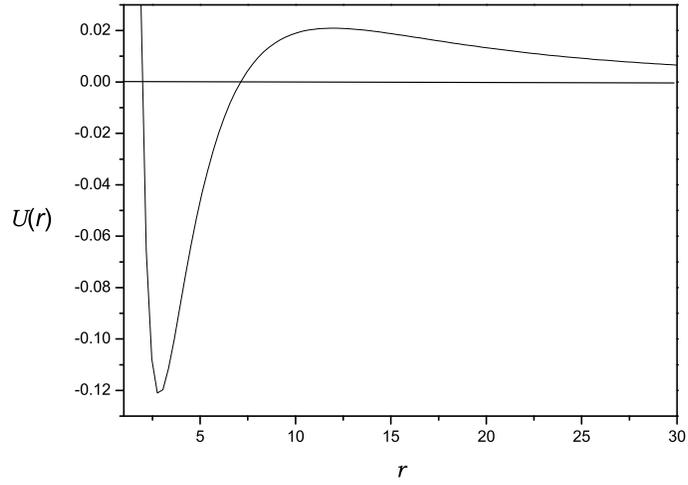}
\end{center}
\caption{$U(r)$ for $l \ne 0$ when $\alpha \approx \alpha_C$.}
\label{fig8}
\end{figure}

\begin{figure} [htbp]
\begin{center}
\includegraphics[totalheight=3.0in]{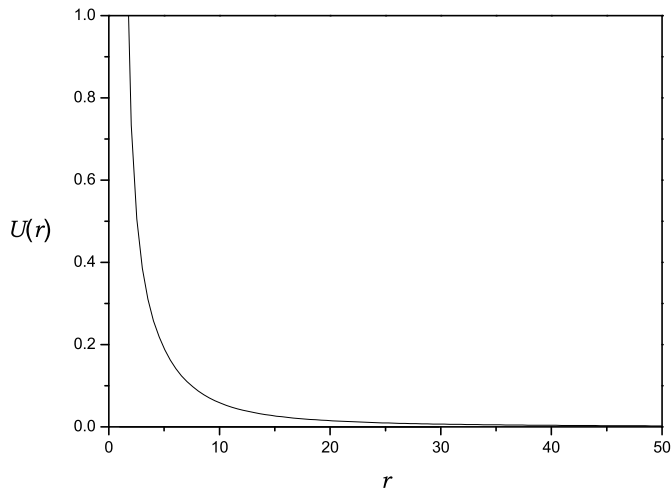}
\end{center}
\caption{$U(r)$ for $l \ne 0$ when $\alpha >> \alpha_C$.}
\label{fig9}
\end{figure}

The second criterion is easy to understand for the $l=0$ case. In such a case,
$U(r)=V(r) < 0$.
If the eigenvalue $E \ge 0$, then $q(r)$ in Eq. (\ref{eq4}) becomes non-negative.
According to QM, we know the state is unbounded.
When $l \ne 0$, special care must be taken.
Figures \ref{fig7}-\ref{fig9} plot $U(r)$ for $l \ne 0$ and
various $\alpha$.
When $\alpha << \alpha_C$, as shown in Fig. \ref{fig7}, there is a well in $U(r)$,
so that some bound states with $E <0$ might exist;
Of course, a state with $E \ge 0$
is unbounded  and
the particle can
travel to arbitrary large distance,
because $q(r)$ is non-negative in the right hand side of the well.
While $\alpha >> \alpha_C$,
as shown in Fig. \ref{fig9}, the effective interaction is repulsive,
and no bound state could exist.

When $\alpha$ is equal to or just above $\alpha_C$, at which
$U(r)$ behaves as Fig. \ref{fig8}, the second term in Eq.
(\ref{eq3}) dominates at intermediate and larger $r$. Is the state
with eigenvalue $E = 0+$ bounded? Without  sufficiently large
$r_{max}$, one would conclude that the state with $E=0+$ to be
bounded. However, with  sufficiently large $r_{max}$, $q(r)$
becomes non-negative at large $r$, and the particle can penetrate
the well and goes to $r \to \infty$ through tunnelling. This means
the state with $E = 0+$ is actually not a bound state.

\begin{figure} [htbp]
\begin{center}
\includegraphics[totalheight=3.0in]{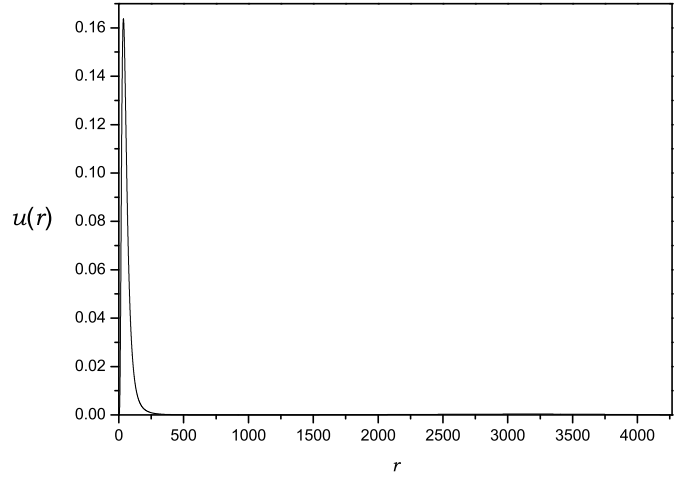}
\end{center}
\caption{Reduced wave function for the ground state with $l=4$,
$\lambda=1$, $\alpha=0.03135$ and $r_{max} < 4267$.} \label{fig10}
\end{figure}

\begin{figure} [htbp]
\begin{center}
\includegraphics[totalheight=3.0in]{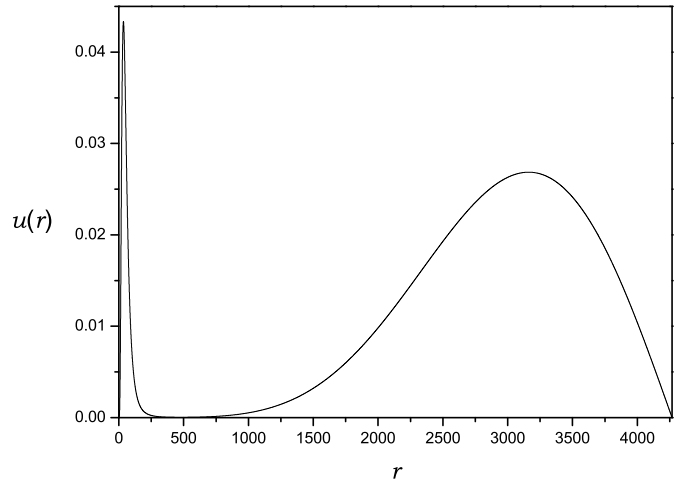}
\end{center}
\caption{Same as  Fig. \ref{fig10}, but with $r_{max}=4267$.}
\label{fig11}
\end{figure}

One should look into the wave function in great detail. Let us
take $l=4$, $\lambda=1$, and $\alpha=0.03135$ as an example.
$U(r)$ behaves like Fig. \ref{fig8}. When $r_{max} \in
[300,4267)$, the system seems to have a bounded ground state with
$E \approx 1.84\times10^{-6}$. The reduced wave function of the
ground state is shown in Fig. \ref{fig10}; The wave function and
energy are stable within this range of $r_{max}$. However,  when
$r_{max} \ge 4267$, as shown in Fig. \ref{fig11}, a peak in $u(r)$
at large $r$ appears. The location of this peak goes to $r \to
\infty$ as increasing $r_{max}$, at the same time, the energy of
the state also tends to zero. Therefore, there is no ground state
for $l=4$, $\lambda=1$, and $\alpha=0.03135$.

\begin{figure} [htbp]
\begin{center}
\includegraphics[totalheight=3.0in]{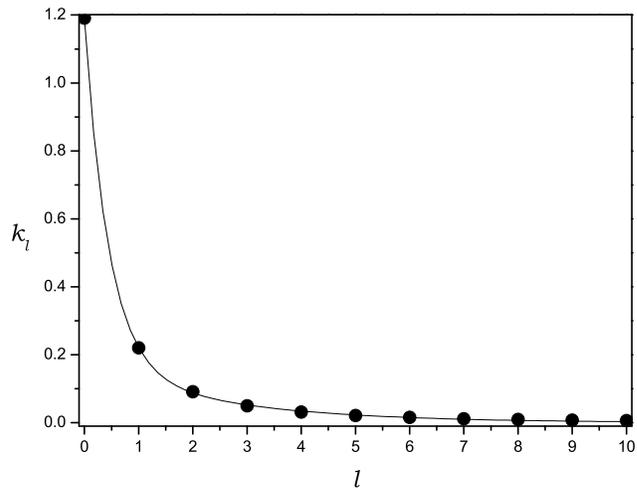}
\end{center}
\caption{$k_{l}$ as a function of $l$.}
\label{fig12}
\end{figure}

Table \ref{tab4} lists the value of $\alpha_C$ for various $l$,
where $k_{l}$ is defined by
\begin{eqnarray}
k_{l}={\alpha_{C} (l) \over \lambda} .
\end{eqnarray}
Figure \ref{fig12} shows $k_l$ as a function of $l$. We can fit the data by
\begin{eqnarray}
k_l=A_{1} \exp(-{l\over B_{1}})+A_{2} \exp(-{l\over B_{2}}) ,
\label{eq12}
\end{eqnarray}
with the fitting parameters given in Tab. \ref{tab5}. For small
$l$, the first term is dominant because $A_1 > A_2$. While for
larger $l$, the second term is dominant, because $B_1 < B_2$. This
implies that $\alpha_C$ decays exponentially for large $l$.

\begin{table}
\begin{center}
\begin{tabular}{|c|c|c|c|c|c|c|}\hline
  $l$ & 0& 1 & 2 & 3 & 4 & 5 \\ \hline
  $k_{l}$ & 1.1906 &
  0.2202 & 0.0913 & 0.0498  & 0.0313 & 0.0215 \\ \hline
  $l$ & 6 & 7 & 8 & 9 & 10 &  \\ \hline
  $k_{l}$ & 0.0157 & 0.0119 & 0.0094 & 0.0076 & 0.0063 &
  \\  \hline
\end{tabular}
\end{center}
\caption{Dependence of $k_{l}$ on $l$.}
\label{tab4}
\end{table}

\begin{table}
\begin{center}
\begin{tabular}{|c|c|c|c|c|}\hline
  fitting coefficient & $A_{1}$& $B_{1}$ & $A_{2}$ & $B_{2}$ \\ \hline
   data & 1.020 & 0.443 & 0.170 & 2.490  \\  \hline
  error  & 0.018 & 0.014 & 0.017 & 0.180  \\ \hline
\end{tabular}
\end{center}
\caption{Fitting parameters in Eq. (\ref{eq12}).} \label{tab5}
\end{table}

\section{Discussions}
\label{SEC5}

In the preceding sections, we have systematically investigated the bound states and critical phenomena
of the Yukawa potential, by means of different numerical algorithms.
The results for the critical screening $\alpha_C$ are consistent with the literature.
We also obtain a new relation between $\alpha_C$ and the angular momentum $l$.

When judging whether a state is bounded,
one must carefully
study the effects of the boundary $r_{max}$  on the wave function and energy.
We believe such an experience might be very useful for numerical study of quantum theory,
in particular at criticality.

While the Runge-Kutta/Numerov algorithm works better in 1D QM, it does not work
in quantum field theory.
In this case, the Monte Carlo methods play an important role.
The Monte Carlo Hamiltonian method has been tested as a useful method for computing the
wave function and spectrum beyond the ground state, at least in QM and the scalar lattice field theory.
We believe that the extension to QCD will allow to explore more useful information
on strong interactions.

\vskip 1.0cm

\noindent
{\bf Acknowledgments}

We thank K Schmidt for useful discussions.

\end{document}